# UX Tri:
# PERIOD CHANGE DISCOVERED
# AND UNCHANGED STRONG BLAZHKO EFFECT CONFIRMED

Herbert Achterberg[1] and Dieter Husar[2,3]

[1] present adress: Dr. H. Achterberg, Liegnitzer Strasse 12, D-22850 Norderstedt (Germany)
[2] Himmelsmoor Private Observatory Hamburg, Himmelsmoor 18, D-22397 Hamburg (Germany)
[3] present adress: Dr. D. Husar, c/o EAT SA, Rue du Séminaire 20A, B-5000 Namur (Belgium);
  email : husar.d (at) gmx.de

**Abstract:**
*New CCD observations of the RRab star UX Tri after JD = 2452234 show a change of the main pulsation period. The period is now* $P = 0.4669046 \pm 0.0000006$ *[d]. The change is* $\Delta P = -1.72 \times 10^{-5}$ *[d] if compared with the value we previously published [1]. A strong Blazhko effect with unchanged Blazhko period* $P_B = 43.7 \pm 0.1$ *[d] is confirmed by this paper. Light curve particularities around Blazhko phase* $\Phi_B = 0.0$ *are discussed in detail. In one observation (JD 2453617) related to Blazhko phase* $\Phi_B = 0.12$ *an exceptionally pronounced "bump" of* $\Delta m > -0.1$ *magnitudes was observed at pulsation phase* $\varphi = 0.71$.

## 1) Introduction

UX Tri was classified as RRab type with a pulsation period of $P = 0.466917$ [d] by Meinunger (1986) after it has been suspected of variability by Morgenroth (1935). More details on the available historical data may be found in our prior IBVS publication of 2001 [1] where we reported our discovery of a Blazhko effect in UX Tri.

## 2) New CCD observations since the publication IBVS 5210 (JD > 2452234)

Since the publication of our IBVS paper on UX Tri in 2001 47 new CCD observations of maxima from this star were made by us and some other observers. Together with the old already known data we have used the results of these new observations to analyse the behaviour of UX Tri again, especially the strong Blazhko effect of this star.

From the whole 47 new observations of maxima since 2001, 22 are obtained from the GEOS RR Lyrae database [2], 14 being made by H. Achterberg and 8 by other persons. 25 new maximum timings, which were all been derived by us, have not been published until now. These are represented in Table 1 of this paper. For one result we acknowledge the help of T. Vanmunster [3]. The mentioned data from the GEOS RR Lyrae database shall not be repeated in this paper here.

As reference and/or comparison stars we used GSC 2294-1202 (GSC: V=11.7 mag; USNO A2.0: R = 11.6 mag), GSC 2294-0900 (GSC: V = 11.6 mag; USNO A2.0: R = 11.6 mag) and GSC 2294-1999 (GSC: V = 11.0 mag; USNO A2.0: R = 11.1 mag). For the maximum timings we used the determination of the maximum of a polynomial and/or in some cases Pogson's method. The systematic errors due to different timing methods is contained in the estimated errors for the maximum timings given in Table 1.

From some observations one of our comparison stars (GSC 2294-0900) can be suspected to be a low amplitude delta Scuti variable. This has however to be studied and confirmed using a photometric set-up with a precision better than 1%.

| JD hel. -2400000 | ± * [d] | (O-C) [d] | $\Phi_B$ | rem. | instr. |
|---|---|---|---|---|---|
| 52250.4183 | 0.0023 | -0.0130 | 0.84 | HSR #16 | cu |
| 52257.3982 | 0.0018 | -0.0366 | 1.00 | HSR #17 | cu |
| 53221.5660 | 0.0250 | -0.0268 | 0.06 | HSR #19 | cu |
| 53272.4951 | 0.0035 | 0.0097 | 0.22 | HSR #20 | f |
| 53285.5619 | 0.0028 | 0.0032 | 0.52 | ATB #53 | a |
| 53291.6246 | 0.0021 | -0.0039 | 0.66 | ATB #54 | a |
| 53316.3906 | 0.0042 | 0.0162 | 0.23 | HSR #24 | cu |
| 53317.3223 | 0.0021 | 0.0140 | 0.25 | HSR #25 | cu |
| 53318.2571 | 0.0021 | 0.0150 | 0.27 | HSR #26 | cu |
| 53321.5262 | 0.0028 | 0.0158 | 0.35 | HSR #27 and VMR #01 | cu & g |
| 53323.3871 | 0.0017 | 0.0091 | 0.39 | HSR #28 | cu |
| 53350.4164 | 0.0050 | -0.0421 | 0.01 | ATB #55 (first max.) | a |
| 53350.4435 | 0.0050 | -0.0150 | 0.01 | ATB #55 (second max.) | a |
| 53387.3198 | 0.0024 | -0.0241 | 0.85 | ATB #56 | a |
| 53394.3347 | 0.0083 | -0.0128 | 0.01 | ATB #58 | a |
| 53408.3683 | 0.0027 | 0.0137 | 0.33 | ATB #59 | a |
| 53617.5488 | 0.0012 | 0.0209 | 0.12 | HSR #29 | cu |
| 53619.4175 | 0.0035 | 0.0220 | 0.16 | HSR #31 | cu |
| 53653.4541 | 0.0015 | -0.0255 | 0.94 | HSR #32 | h |
| 53654.3815 | 0.0024 | -0.0319 | 0.96 | HSR #33 | h |
| 53658.6080 | 0.0070 | -0.0075 | 0.06 | HSR #34 | h |
| 53659.5706 | 0.0056 | 0.0213 | 0.08 | ATB #60 | a |
| 53673.5742 | 0.0016 | 0.0177 | 0.40 | HSR #35 | cu |
| 53674.5086 | 0.0012 | 0.0183 | 0.42 | HSR #36 | cu |
| 53701.5590 | 0.0041 | -0.0117 | 0.04 | HSR #37 | h |
| 53702.5190 | 0.0070 | 0.0145 | 0.06 | HSR #38 | h |

**Table 1:** Unpublished observed times of maxima from CCD observations of the authors. The (O-C)-values are calculated with the elements given in this paper (see next page).

* estimated errors of maximum timings

remarks (rem.): observers ATB = Achterberg, HSR = Husar, VMR = Vanmunster
(the observations are numbered consecutively #NN for each observer)

instrumentation (instr.):
a  0.2 m S.C. Refl. (f = 1100 mm); CCD camera SBIG ST-6 (chip: TI TC241); unfiltered
cu  0.4 m S.C. Refl. (f = 2750 mm); CCD camera SBIG ST-8E (chip: KAF1602E); unfiltered
f  0.2 m S.C. Refl. (f = 950 mm); CCD camera SBIG ST-7 (chip: KAF0400); IR cut-off filter
g  0.35 m S.C. Refl. (f = 2200 mm); CCD camera SBIG ST-7 (chip: KAF0400); unfiltered
h  0.6 m Cass. Refl. (f = 8900 mm); CCD camera: Starlight Express SXV M25C; unfiltered

## 3) Period change and new instantaneous elements for JD > 2452234

When introducing the newly derived maxima in an *(O-C)* graph it is obvious that since 2001 the measured values become more and more negative. It is obvious that a change in the pulsation period has occurred around JD 2452000. In this chapter we derive a new period, the deviation and the rate of the period change.

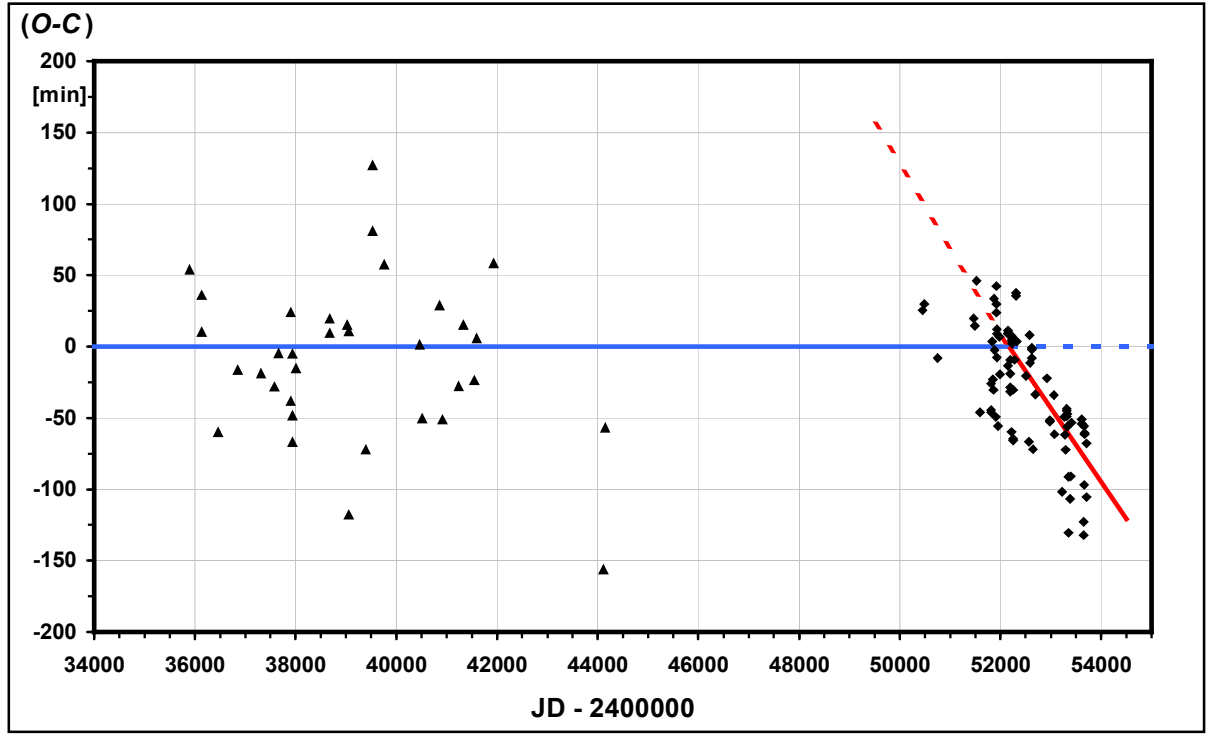

**Fig.1:** *(O-C)* diagram of UX Tri
  all available data: photographic (triangles) and CCD observations (diamonds);
  *(O-C)* values are not corrected for shift due to the Blazhko effect;
  *(O-C)* values are calculated with the elements from [1]: $E_0 = 2450753.488$ and $P = 0.4669218$ [d];
  blue line: target value for the elements from [1];
  red line: target value for the new improved elements from: $E_0 = 2452233.6227$ and $P = 0.4669046$ [d]

In the *(O-C)* diagram of Fig.1 the old photographic observations are plotted as well as all reliable CCD measurements of UX Tri. As the 83 available CCD observations cover only a relatively short time interval, the determination of the pulsation period is influenced by the *(O-C)* scatter caused by the Blazhko effect and a non-homogeneous distribution of the observed maxima over the Blazhko phase. Also the determination of the Blazhko period may be influenced if the pulsation period is not exactly determined. In order to eliminate this interdependences we use an iterative process which we describe in an appendix (chapter 9) in more detail.

Taking into account that the *(O-C)* values are influenced by the *(O-C)* shift caused by the Blazhko effect we eliminated this shift using the known Blazhko period from [1], thus obtaining corrected *(O-C)* values which were then used for further analysis.

From the analysis of the corrected *(O-C)* data with JD > 2452234 an instantaneous period has been calculated (after the above mentioned process):

$$HJD(Max) = 2452233.6227 + 0.4669046 \text{ [d]} \times E$$
$$\pm 0.0011 \pm 0.0000006 \text{ [d]}$$

As the published period in [1] was $P_{IBVS} = 0.4669218 \pm 0.0000003$ this means that there was a change in the period by $\Delta P = -1.72 \times 10^{-5}$ [d]. This change seems to have occurred within a rather short time of < 3 years. This would result in a rate of change of $\Delta P/\Delta t > \approx -2 \times 10^{-8}$.

We have also tried to approximate this period change between JD $\approx$ 2450000 and JD $\approx$ 2454000 by a linear period change $\Delta P/\Delta E \approx -3.9 \times 10^{-9}$ [d], which corresponds to $\Delta P/\Delta t \approx -8.3 \times 10^{-9}$.

Even the precision of the corrected *(O-C)* values are not good enough to decide between these two possibilities, as is shown in Fig.2. On the other hand the two results are quite close and show at least the order of the change, which is almost of the same order of magnitude as derived for RR Lyrae stars with known period changes in Omega Centauri published by Jurcsik et al. in 2001 [4].

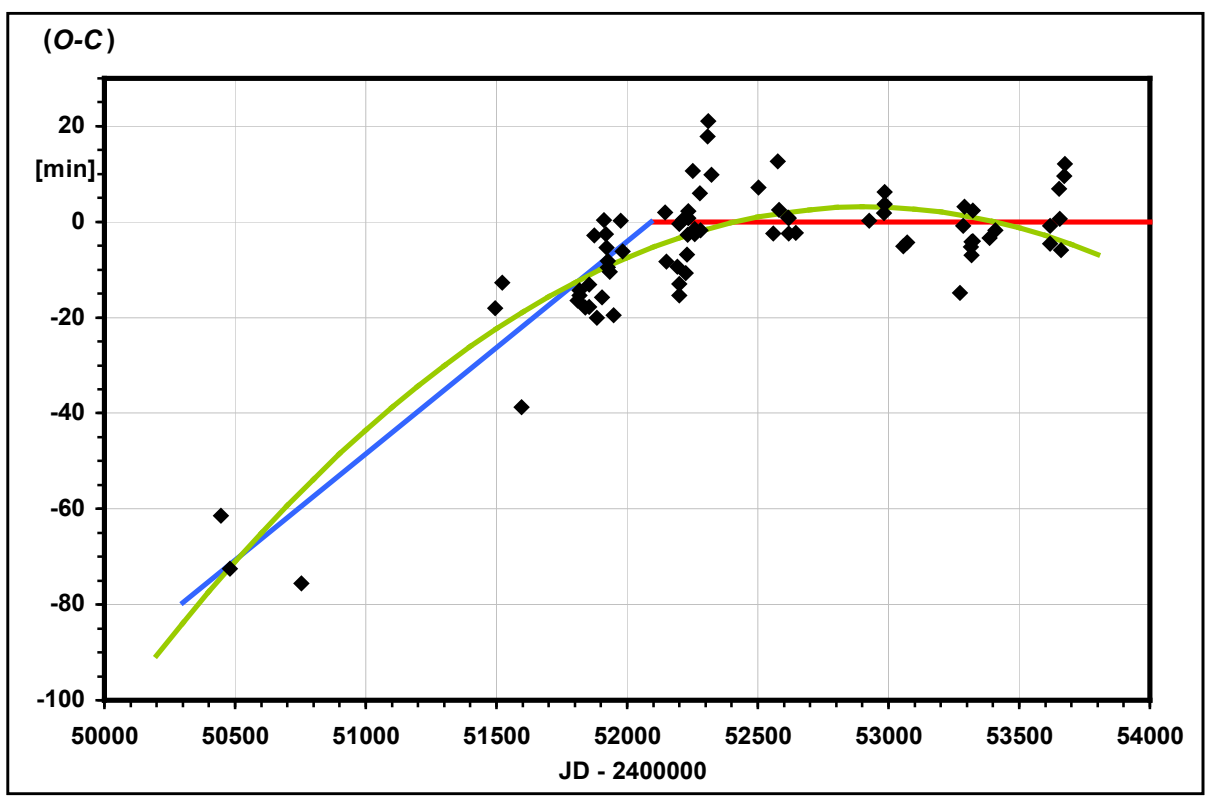

**Fig.2:** *(O-C)* diagram with *(O-C)* values corrected for Blazhko effect:

Only CCD observations between JD 2450466 and JD 2453703 and Blazhko phase > 0.08 are shown (these observations are located on the descending part of the curve in the Blazhko diagram, see Fig. 4)

The *(O-C)* values are corrected for the Blazhko effect ($E_{B0}$ = 2451471; $P_B$ = 43.70 [d]) and calculated with the new elements for observations with JD > 2452234;

red line:
target values for new elements from observations with JD > 2452234: $E_0$ = 2452233.6227, $P$ = 0.4669046 [d];
blue line:
target values for linear elements for observations JD < 2452000: $E_0$ = 2450753.4939, $P$ = 0.46691868 [d];
green parabola:
target values for quadratic elements for observations between JD 2450466 und JD 2453703:
$E_0$ = 2452233.6208, $P$ = 0.46691021 [d], k2 = -1.031 x $10^{-9}$

$E_0$ values are calculated with the condition that the integral over the Blazhko curve was in all cases zero (see Appendix)

#### 4) Determination of the Blazhko period for maxima JD > 2452234

For determination of the Blazhko period $P_B$ we both analysed the *(O-C)* scatter and the changes of the brightness in maximum light as well.

The value of the Blazhko period was determined independently with periodograms from a self written computer program and with the programs 'Period04' [5] and 'Peranso' [6]. An example for a periodogram is shown in Fig.3.

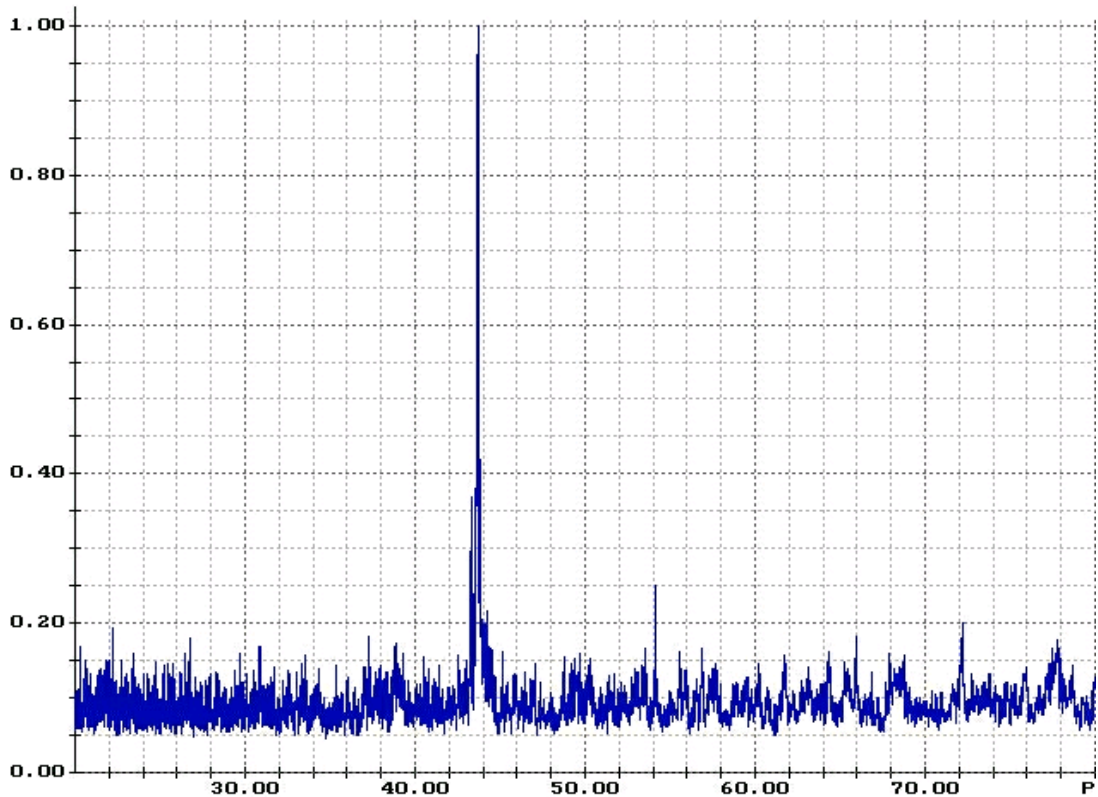

**Fig.3:** Example for a periodogram of the *(O-C)* values with JD > 2452234 calculated with a self written program which is working with the phase-dispersion-minimization (PDM) method. The *(O-C)* values used for the periodogram were calculated with the new derived elements. The values of the *x*-coordinate are in days, in *y*-coordinate relative values are given. The main spectral line in the periodogram is situated at 43.70 [d].

With these methods we determine the Blazhko period $P_{B,T}$ from the *(O-C)* values consistently to be:

$$P_{B,T} = 43.7 \pm 0.1 \text{ [d]}$$

The behaviour of UX Tri within the Blazhko period can be shown clearly, presenting the measured data in a so called Blazhko diagram in which the *(O-C)* values or the brightness of maxima are plotted against the Blazhko phase $\Phi_B$. Blazhko phase $\Phi_B = 0$ is determined by a reference epoch $E_{B0}$, which we have chosen to be the same as in our IBVS publication [1]: $E_{B0}$ = JD 2451471.

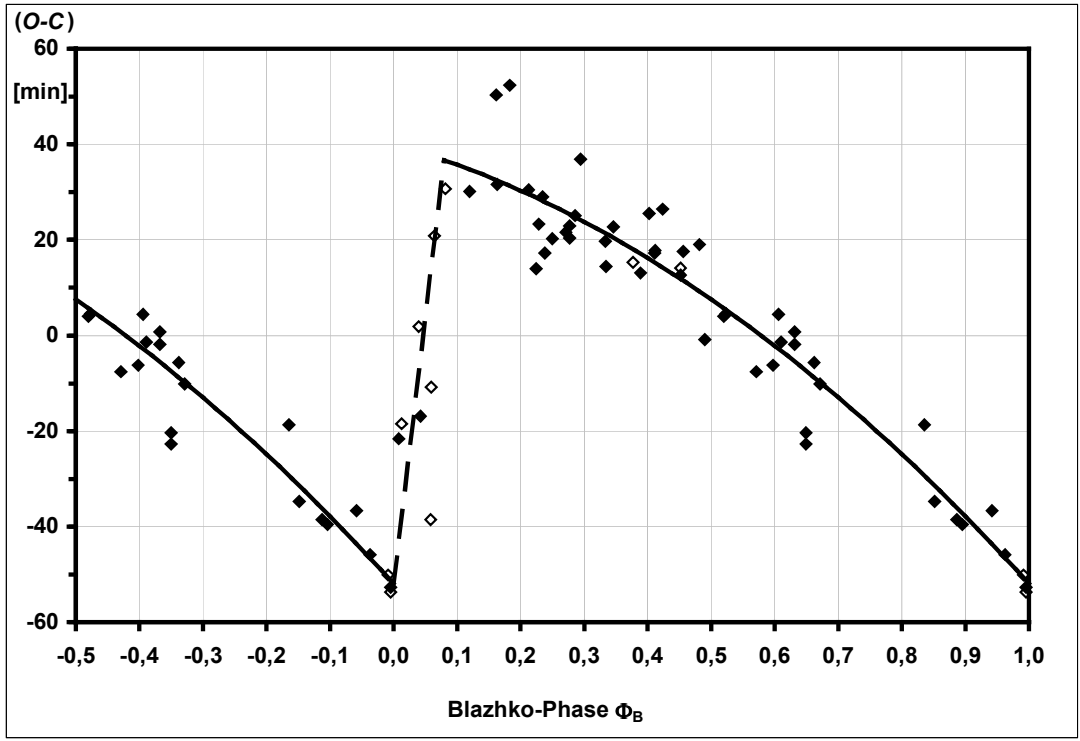

**Fig.4:** Blazhko diagram for the *(O-C)* values of all CCD observations since JD = 2452234.

Blazhko elements used for the diagram are $E_{B0}$ = 2451471, $P_B$ = 43.70 [d].
*(O-C)* values are calculated with the new derived elements.
Open symbols signify uncertain times of maxima.
Solid line: as Blazhko curve we apply a quadratic regression curve to represent the *(O-C)* changes in the phase range from $\Phi_B$ = 0.08 to 1.00.
Dashed line: as Blazhko curve we apply a linear approximation of the steep ascent of the *(O-C)* values.
Both given approximation curves yield the values for the elimination of the Blazhko effect in *(O-C)*.

The Blazhko diagram in Fig.4 for *(O-C)* values with JD > 2452234 shows a fast change of the *(O-C)* values in the range of the Blazhko phase between $\Phi_B$ ≈ 0.0 and $\Phi_B$ ≈ 0.08 which can approximately represented by a straight line (dashed in Fig. 4). The light curves in this Blazhko phase range have rather flat maxima, so that the errors of the measured maximum timings in these cases are often large. This is the main reason why the *(O-C)* values in the area of the steeply ascending slope of the Blazhko curve in Fig. 4 are frequently uncertain. For the rest of the Blazhko period ($\Phi_B$ ≈ 0.08 to $\Phi_B$ ≈ 1.0) the *(O-C)* values gradually decline on an average. This mean slope is represented in Fig. 4 by a quadratic regression curve (solid line) which deviates quite clearly from linearity. The solid and the dashed curves in Fig. 4, here called "Blazhko curve", can also be used to deliver the times of maxima from the time shifts caused by the Blazhko effect.

It should be remarked that the scatter around the curve seems to be quite large compared with the error in the maximum timings. One reason could of course be a change of the pulsation period. We have therefore limited the graph to the observation after JD = 2452234. A search for a systematic periodicity in this behaviour (e.g. a second Blazhko period), was not successful. It is desirable to accumulate much more data in order to succeed with such an analysis. The question if the observed small deviations are regular or irregular remains open for now.

Using the same methods as for the *(O-C)* values we have determined the Blazhko period $P_{B,M}$ from the brightness values in maximum light to be:

$$P_{B,M} = 43.7 \pm 0.1 \text{ [d]}$$

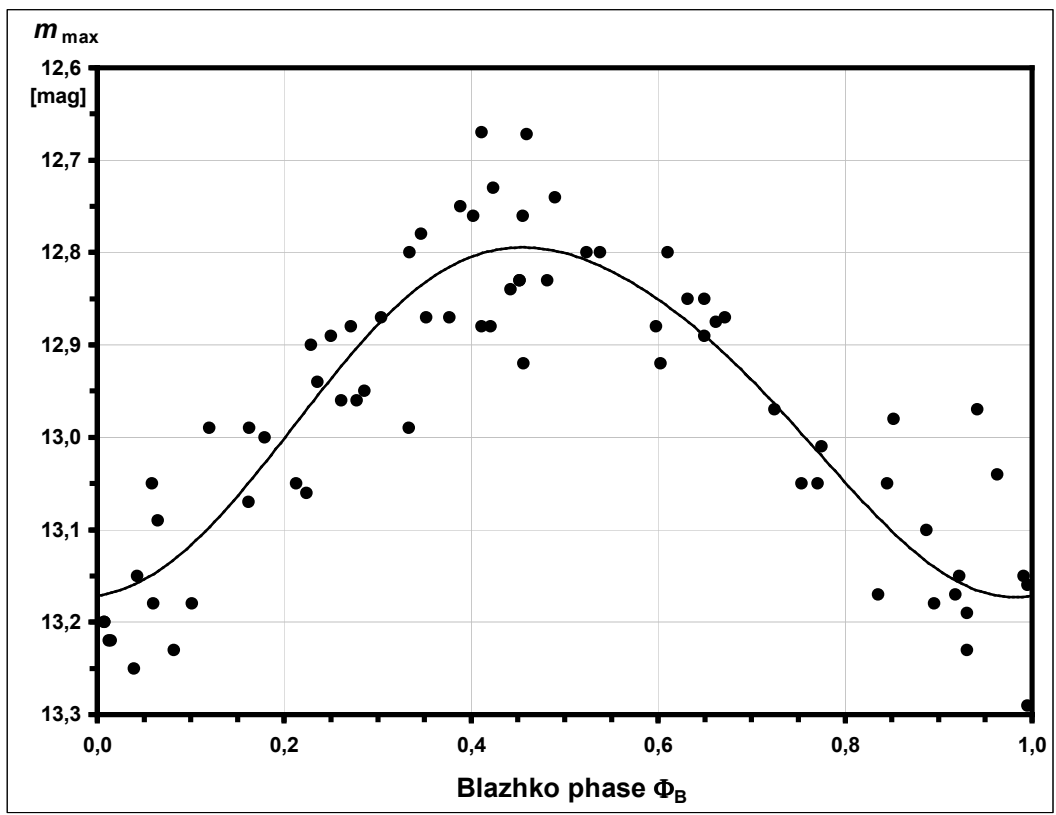

**Fig.5:** Blazhko diagram for the brightness in maximum light of all CCD observations since JD = 2450000 with known brightness values.
Used Blazhko elements are: $E_{B0}$ = 2451471 and $P_B$ = 43.70 [d].

The advantage of this procedure is, that these values are completely independent from the pulsation period, which enables us to use all available CCD observations, despite of the period change. On the other hand the Blazhko diagram for the brightness of maximum (Fig.5) reveals a relatively large spread of the points around the drawn average curve. One reason for this could be systematic errors in the measured brightness values caused by the different equipment used by the observers at the different sites, which we tried however to compensate by an appropriate adjustment.

The first important result is that both methods yield the same period within quite small errors! This is not a trivial result, because it can only be expected, if both the variation of the times of maximum light (scatter in the *(O-C)* values) and the variation of the brightness values at maximum are due to the same physical process.

The second major result is that the Blazhko period remained unchanged within small errors (of ± 0.1 [d]) versus the result from observations from 1999-2001 shown in [1].

## 5) Variations of the light curve of UX Tri during the Blazhko cycle

Already in our first publication of UX Tri we presented the pronounced changes of the light curves over the Blazhko cycle. The fastest light curve changes occur between $\Phi_B \approx 0.9$ and $\Phi_B \approx 0.2$. As this was already evident in 2001 we tried thereafter to give it the highest priority to observe within these limits. It can be seen from Fig. 6 that in fact the changes of the light curve shape are remarkably fast.

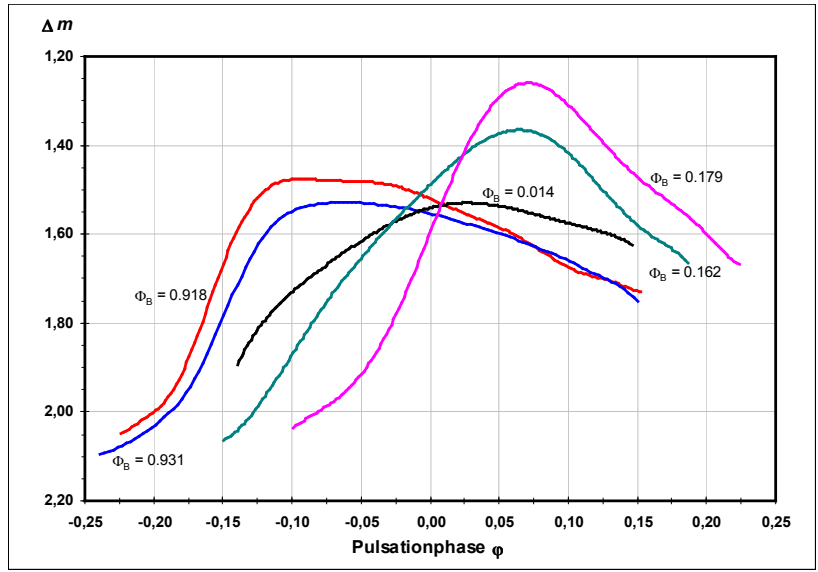

**Fig.6:** In order to show the fast changes due to the Blazhko effect around Blazhko phase
$\mathbf{\Phi_B} \approx 0.0$ we present a few superimposed light curve fits out of this range which are
derived from measured data

We also want to point to the fact that noticeable particularities or irregularities were found in
some of the individual light curves:

α) Around Blazhko phase $\mathbf{\Phi_B} \approx 0.0$ the light curve appears to become very flat (see Fig. 6).
Some very precise light curves show even a slight indication for a double maximum or light
fluctuations during maximum light. A "mean light curve" which has been calculated from
three observations on JD = 2453658 (HSR34; $\mathbf{\Phi_B}$ = 0.060), JD = 2453701 (HSR37; $\mathbf{\Phi_B}$ =
0.043) and JD = 2453702 (HSR38; $\mathbf{\Phi_B}$ = 0.064) by binning the magnitude values of the same
pulsation phase $\mathbf{\varphi}$ is shown in Fig. 7. Again this shows the importance of further photometric
studies especially in this phase of the Blazhko cycle with improved photometry (e.g. relative
photometric errors <1%).

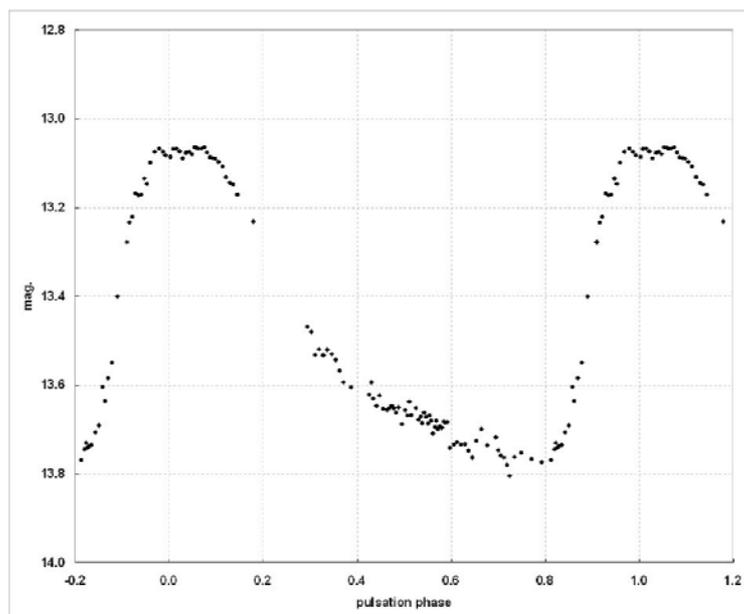

**Fig. 7:** A "mean light curve" which has been calculated from three observations around
Blazhko phase $\mathbf{\Phi_B} \approx 0.0$ showing the flatness in maximum light and some indication
of light fluctuations during the maximum (see text above for details)

β) In one observation (JD 2453617) related to Blazhko phase $\Phi_\mathrm{B}$ = 0.12 an exceptionally pronounced "bump" of $\Delta m$ > -0.1 magnitudes was observed at pulsation phase $\varphi \approx 0.71$. In this case the bump has more or less the aspect of a brightness "flash" (see Fig. 8) which has never been seen so strikingly in our observations. This unfiltered observation of UX Tri was made by D. Husar on 03./04. September 2005 (20:53 – 03:48 UT) during best weather conditions and without moon light. Reference star was GSC 2294-0900. We can locate the effect at JD 2453617.393 (maximum). Fig. 8 shows - except for a slow drift - no unusual behaviour of the comparison stars #2 and #3 during this phase. We also can see a quite usual phenomenon, a broad "hump" at JD 2453617.475, corresponding to pulsation phase $\varphi \approx 0.89$. From literature it is known that so-called "bumps" and "humps" are related to shock waves in RR Lyrae stars, in some cases going hand in hand with exceeding light emission in the U band [7, 8]. As already the observation in unfiltered light shows a significant variation of the light emission occurring in the phase of the pulsation cycle related to the presence of shock waves, it seems probable that the underlying physical processes might be the same. This hypothesis has definitely to be checked by joint filter photometry and high resolution spectroscopy.

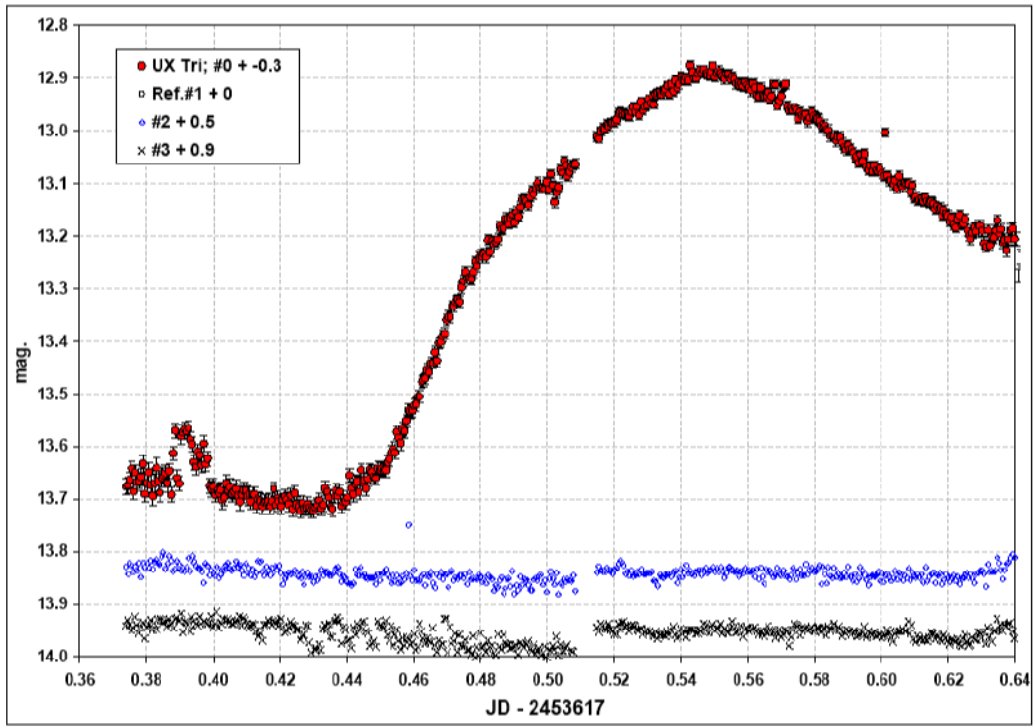

**Fig. 8:** This graph shows the exceptionally pronounced "bump" at JD 2453617.393.

## 6) Final remarks and request for observations to be continued

From the **(O-C)**-Blazhko diagram (Fig.4) and the superimposed and mean light curves (Figures 6 and 7) it is evident that the star shows the most dynamic changes in its Blazhko phase around $\Phi_\mathrm{B}$ = 0.0. We suggest to observe here with the highest priority. As a lot of observed light curve effects have a low brightness amplitude ("flash", magnitude oscillations, double maxima) a relative photometric precision of 0.02 mag or less is very desirable. Filter photometry (e.g. in UBV bands) is preferable in phases of the "bump" and the "hump" in order to check the repeatability of the reported "flash" phenomenon before high resolution spectroscopy may be applied for more detailed studies of the effect.

## 7) Acknowledgements

One of us (D.H.) would like to thank Prof. Dr. Hanns Ruder from the Institut für Theoretische Astrophysik of the University of Tübingen (Germany) for the possibility to use the 0.60-m ROTAT telescope located at the Observatoire de Haute Provence (France) during 6 nights in 2005 for photometry. Further we want to acknowledge the use of the GEOS RR Lyr database [2] as a valuable source for times of maximum data. This research made also use of the SIMBAD data base, operated by the CDS at Strasbourg (France) and related bibliographical services from ADS.

**9) Appendix**

### A METHOD TO DERIVE THE PULSATION PERIOD *P* OF RR LYRAE STARS WITH STRONG BLAZHKO EFFECT WITH IMPROVED ACCURACY

RR Lyrae stars with a strong Blazhko effect show in many cases a large scatter in the *(O-C)* values of the maximum timings, when a linear or quadratic ephemeris formula is used for the calculation. The pulsation period *P* derived with the common least square fit method is in consequence impaired by large errors. As it cannot be expected that the maximum times are homogeneously distributed with respect to the Blazhko phase also systematic errors can occur in the determination of *P*. Often it is not possible to increase precision and accuracy of the derived pulsation period by extending observations over a longer time interval. This may also result from the fact that pulsation periods of RR Lyrae stars - according to experience - are often not constant, so it is necessary to derive *P* from a relatively short time interval of observations.

In these cases we suggest to use a specific method for the derivation of *P* that yields values with satisfying precision and accuracy even when the observations cover a range of only few years. The method, which we will describe here step by step, is essentially based on the idea to correct the observed times of maximum (and with that also the *(O-C)* values) for the fluctuations caused by the Blazhko effect. It is postulated that useful preliminary elements $E_0$ and *P* are available as a first approximation or have been determined with the usual least square fit method.

**1. Determination of the Blazhko period $P_B$**
   If the influence of the Blazhko effect on the *(O-C)* values shall be eliminated the Blazhko period $P_B$ must be known. This period can be determined with suitable period search computer programs from the *(O-C)* values calculated in each case with the best-known elements.

**2. Calculation of the Blazhko phase**
   As soon as a value for the Blazhko period $P_B$ is known every observed time of maximum $O_i$ and therefore also every *(O-C)$_i$* value can be assigned to a Blazhko phase:

$$\Phi_{Bi} = \mathrm{Frac}((O_i - E_{B0})/P_B) \qquad ^1)$$

$E_{B0}$ is an arbitrary but fixed value for the whole investigation. The *(O-C)$_i$* values can be plotted in a so-called Blazhko diagram versus the Blazhko phase $\Phi_{Bi}$. If the used Blazhko period $P_B$ is correct the *(O-C)$_i$* points arrange along a curve with relatively small scatter.

**3. Determination of the Blazhko curve**
   Now it is possible to calculate a best fit to the *(O-C)$_i$* points of the Blazhko diagram as a regression function $y_B = f(\Phi_B)$ (e.g. a second order polynomial). This function we call "Blazhko curve". It can also be composed by sub-functions, which merge into each other.

   Remark:
   For the determination of an accurate value of the pulsation period *P* it is not necessary that the regression function $y_B$ covers the whole Blazhko phase range ($0 < \Phi_B < 1$). It is sufficient when $y_B$ is extended over a rather large part of the phase range from $\Phi_{B1}$ to $\Phi_{B2}$, which includes e.g. about 90% of the whole Blazhko phase range.

---

[1]) Frac(X): the function returns the fractional part of the argument X

## 4. Correcting the observed values $O_i$ and $(O-C)_i$ for the influence of the Blazhko effect

The influence of the Blazhko effect on the measured times of maximum $O_i$ and the $(O-C)_i$ values respectively can now be easily eliminated by subtraction of the value $y_B$ at the corresponding Blazhko phase $\Phi_{Bi}$:

$$O_{icor} = O_i - y_B(\Phi_{Bi}), \qquad (O-C)_{icor} = (O-C)_i - y_B(\Phi_{Bi}).$$

## 5. Calculation of improved elements

Improved elements can now be calculated from the corrected values $(O-C)_{icor}$ by the use of the usual least square fit method. If the pulsation period $P$ is constant in the course of the observations the linear ephemeris formula is appropriate. If however a period $P$ that varies linearly with the number of cycles is assumed then the quadratic formulation of the ephemeris formula has to be applied. In this case the least square fit yields the elements $E_0$ and $P$ and also a value for the coefficient $k_2$ of the quadratic term in the ephemeris formula. If finally a sudden period change occurs in the course of the observations the data have to be treated in two separate groups before and after the period change.

## 6. Iteration

Because the calculated Blazhko curve $y_B$ and possibly the determined Blazhko period $P_B$ as well are dependent on the start value for the pulsation period $P$, for which the correct value just shall be found, it is necessary to derive the final values with an iterative method. Therefore the calculation has to be repeated by starting with step 1 with the calculated corrected values, as new start values until the results of the calculation does not alter more than a given tolerance. E.g. the method described in this paper applied to the data of UX Tri resulted in changes of the pulsation period $P$, which after the fourth iteration step were already smaller than $10^{-8}$ days.

## 7. Determination of the reference epoch $E_0$

Changes in the reference epoch $E_0$ add as constant to the $(O-C)_i$ values, such influencing the vertical position of the Blazhko curve $y_B(\Phi_B)$. In order to define the vertical position of $y_B(\Phi_B)$ it is an appropriate choice to make the mean value of this function equal to zero. That way positive and negative deviations of the $(O-C)_i$ values caused by the Blazhko effect compensate on an average. This may be expressed as condition for $y_B(\Phi_B)$, which must be met by a suitable choice of $E_0$:

$$\int_{\Phi_{B1}}^{\Phi_{B2}} y_B \, d\Phi_B = 0$$

It is sufficient to determine the value of $E_0$ at the end of the iteration procedure because by application of the correct value of $P$ the pulsation period is no more dependent from $E_0$. It shall be mentioned that other definitions of the vertical position of $y_B(\Phi_B)$ are of course possible.

Remark:
After successful determination of the elements the corrected $(O-C)_{icor}$ values lie in the $(O-C)$ diagram in a relatively small band symmetrically to the zero line provided that the pulsation period is constant or changes in a constant manner over the whole observation range.